\documentclass[%
twocolumn,
showpacs,
amsmath,amssymb,
aps,
prb,
]{revtex4-1}

\usepackage{graphicx}
\usepackage{dcolumn}
\usepackage{bm}
\usepackage{natbib}
\usepackage{slashed}
\begin{document}



\title{Enhanced Edelstein effect and interdimensional effects 
in an electron gas with Rashba spin-orbit coupling interface}

\author{A.~C.~Zulkoskey}
 \email[Corresponding author, ]{acz621@mail.usask.ca}

\author{R.~Dick}
 \email{rainer.dick@usask.ca}

\author{K.~Tanaka}
 \email{kat221@campus.usask.ca}

 \affiliation{Department of Physics and Engineering Physics, and Centre for Quantum Topology and Its Applications (quanTA),
University of Saskatchewan, 116 Science Place, Saskatoon, 
Saskatchewan, Canada S7N 5E2}





\date{\today}

\begin{abstract}
We examine the bound-state and free-state contributions to the density of states in a three-dimensional electron gas with a two-dimensional interface with Rashba spin-orbit coupling. Confinement of electrons to the interface is achieved through the inclusion of an attractive potential in the interface. Motivation for our research comes from interest in heterostructure materials that exhibit the Edelstein and inverse Edelstein effects on surfaces or interfaces due to large Rashba spin-orbit coupling. By modifying the Hamiltonian of a three-dimensional free electron gas to include an interface with Rashba spin-orbit coupling and an attractive potential, we are able to calculate the bound-state and free-state wavefunctions and corresponding density of states analytically. We find that 
one of the spin-split energy bands in the interface has an upper bound,
resulting in an enhancement of the Edelstein and inverse Edelstein effect. 
\end{abstract}

\maketitle


\section{\label{sec:level1}Introduction}

The propagation properties of particles (or quasiparticles) affected by the presence of a surface or an interface in a three-dimensional material can be described using low-dimensional quantum mechanics.
Analytic models can be constructed to include extra substructure terms, which affect propagation properties of electrons through a change in the effective mass \cite{rdNR} or confinement in the form of a quantum well \cite{rdqw}. In both cases the Hamiltonian is constructed as a linear superposition of a free three-dimensional electron gas and a low-dimensional substructure contribution describing the effects of a surface or an interface.
The density of states inside the low-dimensional structure that allows for calculation of, e.g., the number of charge carriers and thermal conductivity can be obtained analytically for these types of Hamiltonians and is 
thus a powerful tool for studying the interdimensional properties of electrons in a material with substructure.

The system of a two-dimensional quantum well immersed in a three-dimensional bulk is described by the Hamiltonian, \cite{rdqw}
\begin{equation}
H=\frac{\boldsymbol{p}^2}{2m}-\frac{\hbar^2\kappa}{m}\delta(z-z_0)\,,\label{Eq:qwHam}
\end{equation} 
for a particle of mass $m$.
The quantum well exhibits confining properties through the binding energy, $B=\hbar^2\kappa^2/2m$, with an 
inverse penetration depth $\kappa$. 
The corresponding density of states 
per volume at the location of the quantum-well structure ($z=z_0$) is given as a function of energy $E$ by
\begin{align}
\varrho(E,z_0)=&\kappa \varrho_{d=2}(E+(\hbar^2 \kappa^2/2m))+\varrho_{d=3}(E)\nonumber \\
&\times\Bigg[1-\frac{\hbar\kappa}{\sqrt{2mE}}\arctan\Bigg(\frac{\sqrt{2mE}}{\hbar \kappa}\Bigg)\Bigg],
\end{align}
where
\begin{equation}
\varrho_d(E)=2\Theta(E)\sqrt{\frac{m}{2\pi}}^d\frac{\sqrt{E}^{d-2}}{\Gamma(d/2)\hbar^d}
\end{equation}
is the density of states for a free particle of mass $m$ in $d$ spatial dimensions and the particle is assumed to have spin 1/2 as an electron. Integrating the density of states $\varrho(E,z_0)$ over energy yields the relation between the Fermi energy and the particle density inside the quantum well at zero temperature: \cite{rdqw}
\begin{align}
n&(z_0)\Bigr|_{-B<E_F<0}=\frac{\kappa m}{\pi \hbar^2}\Bigg(E_F+\frac{\hbar^2\kappa^2}{2m}\Bigg)=\kappa n_2\Bigr|_{E_{2,F}=K_{2,F}},\nonumber \\
n&(z_0)\Bigr|_{E_F>0}=\frac{\kappa}{2\pi^2 \hbar^2}\nonumber \\
&\times \Bigg[\hbar \kappa \sqrt{2mE_F}-(\hbar^2\kappa^2 +2mE_F)\arctan\Bigg(\frac{\sqrt{2mE_F}}{\hbar \kappa}\Bigg)\Bigg]\nonumber \\
&+\frac{\kappa m}{\pi \hbar^2}\Bigg(E_F+\frac{\hbar^2\kappa^2}{2m}\Bigg)+\frac{1}{3\pi^2}\Bigg(\frac{\sqrt{2mE_F}}{\hbar}\Bigg)^3,
\end{align}
where 
\begin{align}
n_d=\frac{2}{\hbar^d \Gamma((d+2)/2)}\sqrt{\frac{mE_F}{2\pi}}^d
\end{align}
is the density of particles in $d$ dimensions, and $K_{2,F}=E_F +\hbar^2 \kappa^2/2m$ is the kinetic energy inside the quantum well. The analytic results for the density of states and the particle density inside the quantum well 
smoothly transitions from two-dimensional to three-dimensional behaviour as the inverse penetration 
depth $\kappa$ approaches zero. Both results demonstrate that bound states exist for $E\geq -B$, and that particles confined to the quantum well contribute a two-dimensional density term, made dimensionally correct through 
the factor $\kappa$, reflecting the three-dimensional nature of the system. 

\begin{figure}
  \includegraphics[width=0.85\columnwidth]{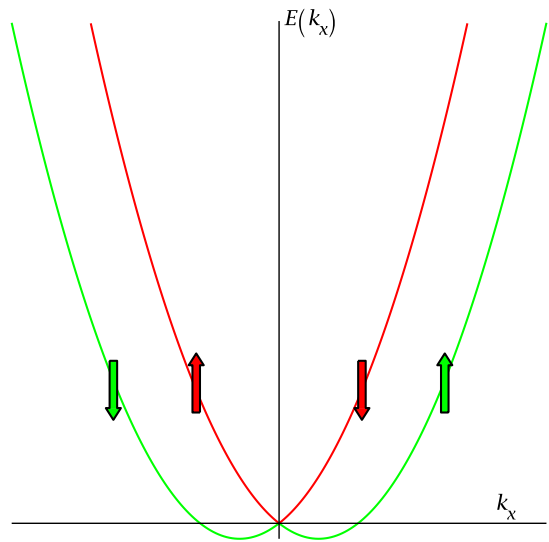}
  \caption{(Colour online) The spin-split dispersion relation for a two-dimensional electron gas with RSOC 
for $k_y=0$ in arbitrary units. The red (green) curve corresponds to $E_+$ ($E_-$) in Eq.~(\ref{Eq:socdisp}) with a minimum energy of $E_{\rm min}=-m\alpha^2/2\hbar^2$. Up and down arrows correspond to $+y$ and $-y$ spin alignment for $k_y=0$. 
  }
  \label{Fig:2Ddispkx}
\end{figure}

Rashba spin-orbit coupling (RSOC) arises 
as a result of bulk inversion asymmetry (BIA), e.g., in the zinc blende structure \cite{bia}, as well as structure inversion asymmetry (SIA) in semiconductors \cite{rash}. The Rashba spin-orbit interaction, first analysed 
for a two-dimensional electron gas,\cite{rash} stems from the nonrelativistic approximation of the Dirac equation \cite{rdb}. The Hamiltonian and dispersion relation for a two-dimensional electron gas including RSOC is given by \cite{rash,rdb}
\begin{align}
&H=\frac{\boldsymbol{p_\parallel}^2}{2m}+\alpha\left[\boldsymbol{\sigma}\times\boldsymbol{p}_\parallel/\hbar\right]\cdot \hat{z}\,,\nonumber \\
&E_\pm(k_\|)=\frac{\hbar^2 k_\parallel^2}{2m}\pm \alpha |\boldsymbol{k}_\parallel|\,,\label{Eq:socdisp}
\end{align}
where $m$ is the effective mass of an electron, $\alpha=e\hbar E_z(z)/4m^2 c^2$ is the Rashba coefficient, $E_z(z)$ is 
an electric field in the direction $\hat{z}$ perpendicular to the electron gas, $\boldsymbol{\sigma}=(\sigma_x,\sigma_y,\sigma_z)$ are the Pauli matrices, $\boldsymbol{k}_\parallel=(k_x,k_y)$ is the two-dimensional wave vector, and $k_\|=|\boldsymbol{k}_\parallel|$. 
RSOC causes momentum and spin to be ``locked in'' such that the $E_+$ and $E_-$ branches in Eq.~(\ref{Eq:socdisp}) have clockwise and counterclockwise winding of spin, respectively, as one goes around the Fermi surface. In Fig.~\ref{Fig:2Ddispkx}, $E_\pm$ are plotted as a function of $k_x$ for $k_y=0$, where spin points in either $+y$ or $-y$ direction.
The density of states per 
unit area per spin, 
\begin{equation}
\varrho(E)=\frac{1}{2\pi}\,\frac{k_\parallel(E)}{|dE/dk_\parallel|}\,,
\label{Eq:bsdisp}
\end{equation}
for $E=E_\pm$ using Eq.~(\ref{Eq:socdisp}) is given by \cite{wink}
\begin{align}
\varrho(E_\pm)&=\frac{m}{2\pi\hbar^2}\Bigg(1\mp \frac{b}{\sqrt{b^2+2Em/\hbar^2}}\Bigg),\quad E\geq 0\,,\nonumber \\
\varrho(E_{-})&=\frac{mb}{\pi\hbar^2\sqrt{b^2+2mE/\hbar^2}},\quad E<0\,,\label{Eq:vh2d}
\end{align}
where $b=m\alpha/\hbar^2$.

Recent efforts have been put forth in the area of spintronics, which utilizes the spin degree of freedom for information storage and processing.\cite{spintronics} In candidate materials 
for spintronics, 
strong RSOC induces novel properties on surfaces or interfaces \cite{indsoc} 
such as the Edelstein effect \cite{edelstein} or the inverse Edelstein effect,
where conversion between charge and spin currents occurs.\cite{indsoc,qmat} 
We study this kind of effects on the density of states for Hamiltonians which are a linear combination of three-dimensional kinetic terms and a two-dimensional RSOC term, in order to model systems with large RSOC on an interface or a surface. Materials which necessitate this description include topological insulators,\cite{topins} interfaces between metallic layers, e.g., Bi/Ag \cite{biag_inverse,biag} or Cu/Bi \cite{cubi} interfaces, and conducting interfaces between $\rm{LaAlO}_3/\rm{SrTiO}_3$ insulating oxide layers.\cite{insoxide} Heterostructures involving metal-oxide interfaces \cite{metaloxide} as well as graphene,\cite{graphene1,graphene2} in which RSOC is enhanced by proximity to, e.g., transition metal dichalcogenides,\cite{indsoc,qmat} also present systems in which RSOC is prominent along an interface or a surface. In this paper we study the low-energy physics on a surface or an interface in such materials.
The remaining sections are laid out as follows. 
In Sec.~\ref{sec:level2} we calculate the bound-state and free-state wavefunctions of our interdimensional model. In Sec.~\ref{sec:level3} the enhancement of the Edelstein effect for electrons bound to the interface is demonstrated. In Sec.~\ref{sec:level4} we present the analytic results for the bound-state and free-state density of states and discuss the interdimensional behaviour. Our findings are summarised in Sec.~\ref{sec:level5}.

\section{\label{sec:level2}Interdimensional effects of electrons with RSOC interface}
\begin{figure}
  \includegraphics[scale=0.35]{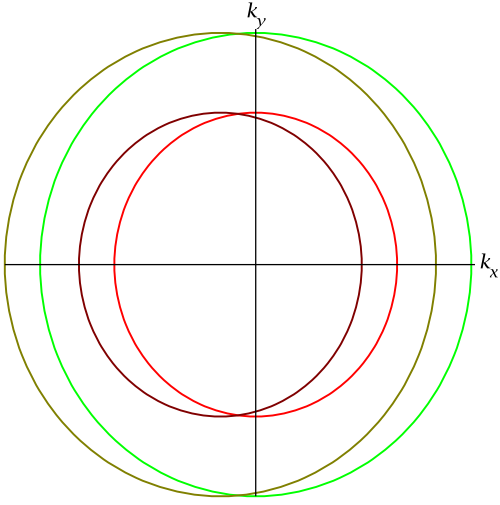}
  \caption{(Colour online) The shifted Fermi surfaces of a two-dimensional electron gas with RSOC due to 
an applied electric field in the $x$ direction.
The red (green) circle corresponds to the $E_+$ ($E_-$) branch for the applied electric field $\bm{E}=0$ and the maroon (olive) circle corresponds to the $E_+$ ($E_-$) branch for $\bm{E}\neq 0$.}
  \label{Fig:2DFC}
\end{figure}
Motivated by materials which exhibit novel features on interfaces or surfaces as a result of RSOC, as well as heterostructures where RSOC in the interface is enhanced by the neighbouring substrate \cite{indsoc,qmat}, we construct a Hamiltonian as a superposition of a three-dimensional free electron gas and a two-dimensional interface with RSOC and an attractive potential at $z=z_0$. We extend the work of Ref.~\onlinecite{rdNR} to include a RSOC term \cite{rash} in the interface, 
\begin{equation}
H_{SO}=\alpha \left[ \boldsymbol{\sigma}\times \boldsymbol{k}_\parallel\right] \cdot \hat{z}\,.\label{Eq:rashterm}
\end{equation}
The construction of the Hamiltonian for electrons subject to spin-orbit coupling in the interface is as follows: We assume that the wavenumber component orthogonal to the interface is small compared to the inverse thickness of the interface $L_\perp^{-1}$, i.e., $|k_\perp L_\perp|\ll1$. This implies that the wavefunction in the direction orthogonal to the interface can be approximated as constant. This formulation yields a second-quantized Hamiltonian,
\begin{align}
H&=\int d^3\boldsymbol{x}\,\frac{\hbar^2}{2m}\boldsymbol{\nabla}\psi^\dagger(\boldsymbol{x})\cdot\boldsymbol{\nabla}\psi(\boldsymbol{x})\nonumber \\
&-\int d^2\boldsymbol{x}_\parallel\Bigg( i\alpha L_\perp \psi^\dagger(\boldsymbol{x}_\parallel,z_0)\left(\boldsymbol{\sigma}_\parallel \times \boldsymbol{\nabla}_\parallel\right) \cdot \hat{z}\psi(\boldsymbol{x}_\parallel,z_0)\nonumber \\
&+V_0\psi^\dagger(\boldsymbol{x}_\parallel,z_0)\cdot\psi(\boldsymbol{x}_\parallel,z_0)\Bigg),\label{Eq:rashHam2ndSOCAP}
\end{align}
where $\boldsymbol{x}=(\boldsymbol{x}_\parallel,z)$ and $\boldsymbol{\sigma}_\parallel= (\sigma_x,\sigma_y)$. The eigenvalues and eigenfunctions of Eq.~(\ref{Eq:rashHam2ndSOCAP}) are separated into states which are bound to the interface $(E<\hbar^2 k_\parallel^2/2m, \kappa>0)$,
\begin{align}
&\psi_{\boldsymbol{k}_\parallel,\kappa_\pm,+}(\boldsymbol{x}_\parallel,z)_\pm=\langle \boldsymbol{x}_\parallel,z|\boldsymbol{k}_\parallel,\kappa,+\rangle_\pm \nonumber \\
&\hspace{.4cm}=\frac{\exp(i\boldsymbol{k}_\parallel\cdot\boldsymbol{x}_\parallel)}{2\pi}\sqrt{\kappa_\pm}\exp(-\kappa_\pm|z-z_0|)\varphi(\boldsymbol{k_\|})_\pm\,, \label{Eq:bswfAP}
\end{align}
\begin{equation}
E_{\pm}=\frac{\hbar^2}{2m}(\boldsymbol{k}_\parallel^2-\kappa_\pm^2)=\frac{\hbar^2 k_\parallel^2}{2m}-\frac{mL_\perp^2}{2\hbar^2}(V_0\pm \alpha k_\|)^2\label{Eq:bsdisp2},
\end{equation}
where $\varphi(\boldsymbol{k_\|})_\pm=\frac{1}{\sqrt{2}}\binom{1}{\pm ik_+/k_\|}$, $k_\pm = k_x\pm ik_y$, and $\kappa_\pm = (m L_\perp/\hbar^2) (V_0\pm \alpha k_\|)$,
and two sets of orthogonal, transversely free states ($k_\perp\geq0$, $E=\hbar^2(\boldsymbol{k}_\parallel^2+k_\perp^2)$) written as even ($+$) and odd ($-$) parity eigenstates,
\begin{align}
&\psi_{\boldsymbol{k}_\parallel,k_\perp,+}(\boldsymbol{x}_\parallel,z)_\pm=\langle \boldsymbol{x}_\parallel,z|\boldsymbol{k}_\parallel,k_\perp,+\rangle_\pm \nonumber \\
=&\frac{\exp(i\boldsymbol{k}_\parallel\cdot\boldsymbol{x}_\parallel)}{2\sqrt{\pi^3}\sqrt{1+(mL_\perp/\hbar^2 k_\perp)^2(V_0 \pm \alpha k_\|)^2}}\Bigg(\cos(k_\perp (z-z_0))\nonumber \\
&-\frac{m L_\perp}{\hbar^2 k_\perp}(V_0\pm \alpha k_\|)
	\sin(k_\perp |z-z_0|)\Bigg)\varphi(\boldsymbol{k_\|})_\pm\,,\label{Eq:fswfAP}
\end{align}
\begin{align}
\psi&_{\boldsymbol{k}_\parallel,k_\perp,-}(\boldsymbol{x}_\parallel,z)=\langle \boldsymbol{x}_\parallel,z|\boldsymbol{k}_\parallel,k_\perp,-\rangle\nonumber \\
&=\frac{\exp(i\boldsymbol{k}_\parallel\cdot\boldsymbol{x}_\parallel)}{2\pi}\frac{1}{\sqrt{\pi}}\sin(k_\perp (z-z_0))\binom{\cos\xi}{e^{i\chi}\sin\xi},
\end{align}
where $\chi$ and $\xi$ are arbitrary real numbers. Note that the interface bound states (\ref{Eq:bswfAP}) have even (+) parity.
In order for the ground-state energy to exist as a lower bound in Eq.~(\ref{Eq:bsdisp2}) we require $\eta^2\leq 1$ where $\eta=m\alpha L_\perp/\hbar^2$. Without this restriction $E\to-\infty$ as $k_\|\to\infty$. 
Thus, the Hamiltonian in Eq.~(\ref{Eq:rashHam2ndSOCAP}) yields the dispersion relation,
\begin{align}
E=E(|\boldsymbol{k}_\| |,\kappa_\pm)\equiv E_\pm(|\boldsymbol{k}_\| |)
\end{align}
for bound states $\{|\boldsymbol{k}_\|,\kappa, + \rangle_\pm\}$ with $E<\hbar^2\boldsymbol{k}_\|^2/2m$, and 
\begin{align}
E=E(|\boldsymbol{k}_\| |,k_\perp)
\end{align}
for the unbound states $\{|\boldsymbol{k}_\|,k_\perp, \pm \rangle_\pm\}$ with $E\ge\hbar^2\boldsymbol{k}_\|^2/2m$. 
The corresponding 
density of states is given by 
\cite{rdNR}
\begin{align}
\varrho&(E,\boldsymbol{x})=\sum_{\pm}\int d^2\boldsymbol{k}_\|\Bigg(\delta(E-E_\pm(|\boldsymbol{k}_\| |))|\langle \boldsymbol{x}|\boldsymbol{k}_\|,\kappa\rangle_\pm |^2\nonumber \\
&+\Theta(E-\hbar^2\boldsymbol{k}_\|^2/2m)\Biggr|\frac{\partial k_\perp(E,\boldsymbol{k}_\|)}{\partial E}\Biggr||\langle \boldsymbol{x}|\boldsymbol{k}_\|,k_\perp\rangle_\pm |^2 \Bigg)
\label{Eq:DOSwffsbs},
\end{align}
where the sum over even- and odd-parity unbound states is implicitly assumed, and as such $+$ and $-$ for parity have been removed from the eigenvectors.
Hence,
\begin{align}
&\varrho(E,\boldsymbol{x})=\sum_{\pm}\Bigg(\int^{2\pi}_0 d\theta\, k_\|\Biggr| \frac{\partial k_\|(E,\kappa_\pm)}{\partial E}\Biggr||\langle \boldsymbol{x}|\boldsymbol{k}_\|,\kappa\rangle_\pm |^2 \nonumber \\
 &+\int d^2\boldsymbol{k}_\|\, \Theta(E-\hbar^2\boldsymbol{k}_\|^2/2m)\Biggr|\frac{\partial k_\perp(E,\boldsymbol{k}_\|)}{\partial E}\Biggr||\langle \boldsymbol{x}|\boldsymbol{k}_\|,k_\perp\rangle_{\pm} |^2\Bigg). \label{Eq:DOSwffsbs2}
\end{align}

\section{\label{sec:level3}Bound-state dispersion relation and enhanced Edelstein effect}
In a purely two-dimensional system with RSOC, 
an applied electric field along $+x$ causes electrons to move in the $-x$ direction and populate states with $k_x<0$ at the expense of states with $k_x>0$. Figure~\ref{Fig:2DFC} illustrates the shift of the inner and outer Fermi circles 
due to an applied electric field in the $x$ direction.
An increase in $-y$ and $+y$ spin polarization states for $k_y=0$ creates a net $-y$ spin polarization, as the outer Fermi circle dominates over the inner one with a larger number of states. 
This is the well-known Edelstein effect,\cite{indsoc,edelstein,qmat} where a charge current is converted to an accumulation of spin in the transverse direction. Likewise, the inverse Edelstein effect is the conversion of a spin current to a transverse charge current.\cite{indsoc,qmat,biag_inverse}

\begin{figure}
  \includegraphics[scale=0.35]{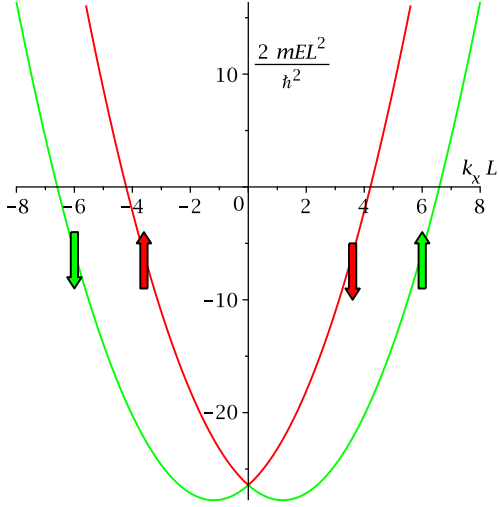}
  \caption{(Colour online) The spin-split bound-state energy $(2mL_\perp^2/\hbar^2)E_\pm$ as a function of $k_xL_\perp$
for $k_y=0$. The red (green) curve corresponds to $E_-$ ($E_+$) in Eq.~(\ref{Eq:bsdisp2}) with a minimum dimensionless energy of $2mE_{\rm min}L_\perp^2/\hbar^2=-\beta^2/(1-\eta^2)$, where $\beta=mV_0L_\perp^2/\hbar^2$ and $\eta=m\alpha L_\perp/\hbar^2$.}
  \label{Fig:bsdispAP1}
\end{figure}
\begin{figure}
  \includegraphics[scale=0.35]{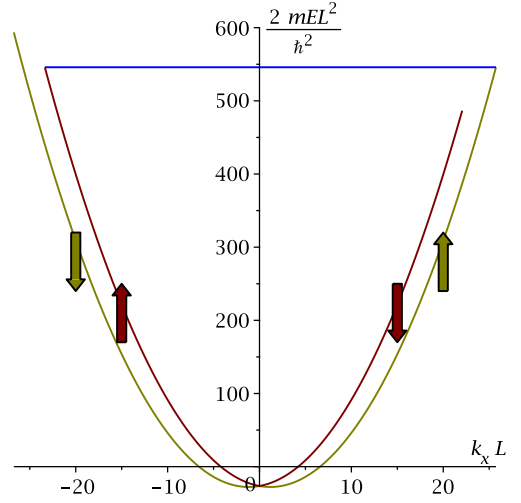}
  \caption{(Colour online) The spin-split bound-state energy $(2mL_\perp^2/\hbar^2)E_\pm$ as a function of $k_xL_\perp$ 
for $k_y=0$. The red (green) curve corresponds to $E_-$ ($E_+$) in Eq.~(\ref{Eq:bsdisp2}). Due to the requirement $\kappa_->0$, $E_-$ is restricted to a maximum dimensionless energy of $(V_0L_\perp/\alpha)^2 = \beta^2/\eta^2$, where $\beta=mV_0L_\perp^2/\hbar^2$ and $\eta=m\alpha L_\perp/\hbar^2$, indicated by the blue line.}
  \label{Fig:bsdispAP2}
\end{figure}
\begin{figure}
  \includegraphics[scale=0.35]{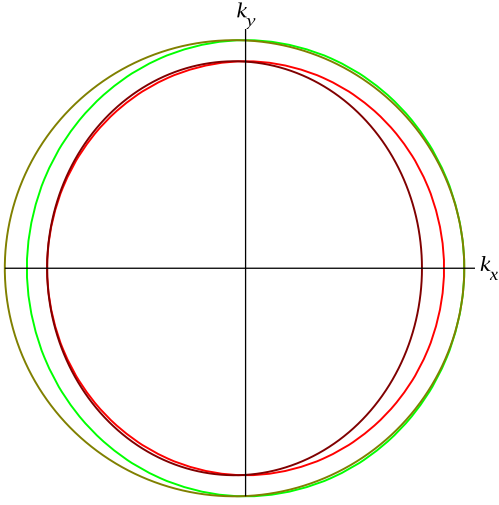}
  \caption{(Colour online) The shifted Fermi surfaces due to the enhanced Edelstein effect. The red and green circles correspond to $E_-$ and $E_+$, respectively, in Eq.~(\ref{Eq:bsdisp2}), while the maroon and olive circles correspond to the situation shown in Fig.~\ref{Fig:bsdispAP2}.}
  \label{Fig:bsFCAP}
\end{figure}

In our three-dimensional model 
we find that the Edelstein effect \cite{edelstein} (and inverse Edelstein effect) is affected through the restriction $\kappa_->0$. 
The spin-split bound-state bands in Eq.~(\ref{Eq:bsdisp2}), made dimensionless with a factor $2mL_\perp^2/\hbar^2$, are plotted as a function of $k_xL_\perp$ for $k_y=0$ in Fig.~\ref{Fig:bsdispAP1}. 
Note that $E_- \ge E_+$ for $k_\| \ge 0$.
The requirement $\kappa_->0$ implies $k_\|<V_0/\alpha$ in order for a bound-state solution to exist. This restricts the energy $E_-$ in Eq.~(\ref{Eq:bsdisp2}) to a maximum value of $E_-=\hbar^2 V_0^2/2m\alpha^2$. Consider the situation of $E_F\lesssim \hbar^2V_0^2/2m\alpha^2$. Applying an electric field 
along the interface 
results in a net spin polarization in the $-y$ direction due to the Edelstein effect. However, the states belonging to the $E_-$ branch can only be populated up to the maximum dimensionless energy of $V_0^2L_\perp^2/\alpha^2$. The contribution to the net spin polarization from the increase in $+y$ spin alignment in the $k_x<0$ region of the $E_-$ branch therefore has an upper bound. The $E_+$ branch has no restriction owing to $\kappa_+>0;\;\forall k_\|$, which leads to a larger net spin polarization in the $-y$ direction compared to the purely two-dimensional model and hence an enhanced Edelstein effect. The dimensionless energy bands $(2mL_\perp^2/\hbar^2)E_\pm$ are plotted as a function of $k_xL_\perp$ for $k_y=0$ in Fig.~\ref{Fig:bsdispAP2} along with the maximum allowed $E_-$ for the bound states. The energy cap on $E_-$ results in shifting of the Fermi circles to elliptical shape, as illustrated in Fig.~\ref{Fig:bsFCAP}.

\section{\label{sec:level4}Bound-state and Free-state density of states at $z=z_0$}
\begin{figure}
  \includegraphics[scale=0.35]{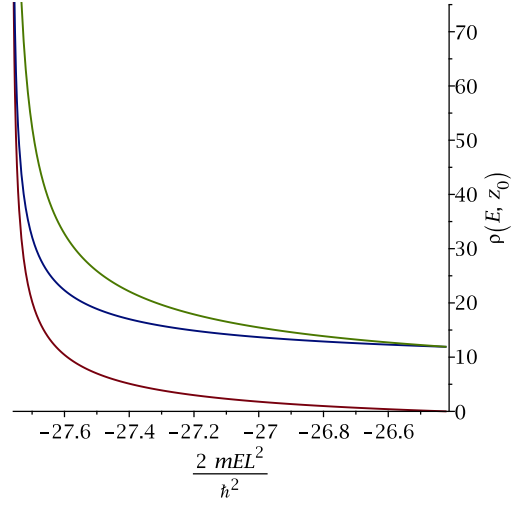}
  \caption{(Colour online) The spin-split and total bound-state density of states at $z=z_0$ for $E_{\rm min}\leq E\leq-mV_0^2L_\perp^2/2\hbar^2$ in units of $m/(2\pi\hbar^2L_\perp)$. The green curve is the total bound-state density of states, while the blue (red) curve is the contribution from the $+$ ($-$) sign choice in the solutions for $k_\|$ in $E_+$.}
  \label{Fig:bsDOSAP_lowE}
\end{figure}

\begin{figure}
  \includegraphics[scale=0.35]{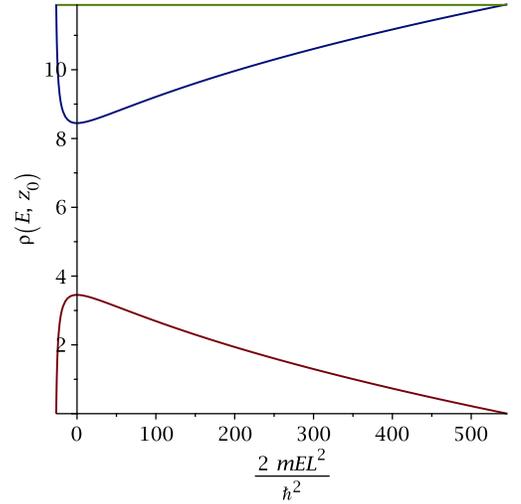}
  \caption{(Colour online) The spin-split and total bound-state density of states at $z=z_0$ for $-mV_0^2L_\perp^2/2\hbar^2\leq E \leq \hbar^2V_0^2/2m\alpha^2$ in units of $m/(2\pi\hbar^2L_\perp)$. The green curve is the total bound-state density of states, while the blue and red curves correspond to, respectively, $\varrho(E_+,z_0)$ and $\varrho(E_-,z_0)$ in Eq.~(\ref{Eq:ssbsDOS}).}
  \label{Fig:bsDOSAP}
\end{figure}

The restriction on the energy range for the $E_-$ branch also impacts the bound-state density of states at $z=z_0$. Substituting Eq.~(\ref{Eq:bswfAP}) in Eq.~(\ref{Eq:DOSwffsbs2}) yields the density of states of the spin-split energy bands in three energy ranges. For $E_{\rm min}\leq E < -mV_0^2L_\perp^2/2\hbar^2$ only the $E_+$ branch contributes to the density of states and is given by
%
\begin{equation}
\varrho(E_+,z_0)=\frac{m^2 \eta\,( V_0^2 L_\perp^2+ f(E,\eta,V_0))}{\pi\hbar^4\sqrt{f(E,\eta,V_0)}(1-\eta^2)^2}\,,\label{Eq:BSDOS_lowE}
\end{equation}
where $E_{\rm min}=-mV_0^2L_\perp^2/2\hbar^2(1-\eta^2)$, $f(E,\eta,V_0)=V_0^2 L_\perp^2+ (2E\hbar^2/m)(1-\eta^2)$ and $\eta=m\alpha L_\perp/\hbar^2$. Equation~(\ref{Eq:BSDOS_lowE}) shows that the density of states for $E_{\rm min}\leq E\leq -mV_0^2L_\perp^2/2\hbar^2$ contains a van Hove singularity at $E=E_{\rm min}$. This singularity occurs at the bottom of the $E_+$ branch, which is analogous to 
the van Hove singularity at $E=-m\alpha^2/2\hbar^2$ in the purely two-dimensional model.\cite{rash}
Shown in Fig.~\ref{Fig:bsDOSAP_lowE} is the density of states in the energy range $E_{\rm min}\leq E \leq -mV_0^2L_\perp^2/2\hbar^2$. The density of states of the spin-split bound states in the energy range $-mV_0^2L_\perp^2/2\hbar^2 \leq E\leq \hbar^2 V_0^2/2m\alpha^2$ is displayed in Fig.~\ref{Fig:bsDOSAP} and is given by
\begin{align}
&\varrho(E_\pm,z_0)=\frac{mL_\perp}{2\pi\hbar^2}\Bigg[\Bigg(\frac{\pm mV_0  \alpha L_\perp^2/\hbar^2+\sqrt{f(E,\eta,V_0)}}{(\sqrt{f(E,\eta,V_0)})(\hbar^2/m)(1-\eta^2)}\Bigg)\nonumber \\
&\times\Bigg(V_0\pm\alpha\frac{\pm mV_0  \alpha L_\perp^2/\hbar^2+\sqrt{f(E,\eta,V_0)}}{(\hbar^2/m)(1-\eta^2)}\Bigg)\Bigg]\nonumber \\
&\times\Theta(E+mV_0^2L_\perp^2/2\hbar^2)\label{Eq:ssbsDOS}.
\end{align}
The total density of states corresponding to Eq.~(\ref{Eq:ssbsDOS}) is
\begin{equation}
\varrho(E,z_0)=\frac{\varrho_{d=2}(K_2)\beta}{L_\perp(1-\eta^2)^2}\,, \label{Eq:bsDOS}
\end{equation}
where $K_2=E+mV_0^2L_\perp^2/2\hbar^2$ is the kinetic energy of electrons whose wavefunctions are exponentially suppressed perpendicular to the interface, and $\beta=mV_0L_\perp^2/\hbar^2$.
Equation~(\ref{Eq:bsDOS}) demonstrates that the total density of states in the energy range where both spin-split bands contribute is proportional to the free two-dimensional density of states scaled by the interface thickness $L_\perp$, to reflect the three-dimensional nature of the system. For $E>\hbar^2V_0^2/2m\alpha^2$ the $E_-$ branch no longer contributes and the bound-state density of states is given by $\varrho(E_+,z_0)$ in Eq.~(\ref{Eq:ssbsDOS}). This is shown in Fig.~\ref{Fig:bsDOSAP}.
The free-state density of states is calculated by inserting Eq.~(\ref{Eq:fswfAP}) into Eq.~(\ref{Eq:DOSwffsbs2}),
\begin{align}
&\varrho(E,z_0)=\frac{m\Theta(E)}{2\pi^2\hbar^2L_\perp}\int^{\sqrt{2mEL_\perp^2}/\hbar}_0 dx \nonumber \\
&\Bigg(\frac{x\sqrt{2mEL_\perp^2/\hbar^2-x^2}}{2mEL_\perp^2/\hbar^2-x^2+(m L_\perp/\hbar^2)^2(-V_0 +\alpha x)^2}\nonumber \\
&+\frac{x\sqrt{2mEL_\perp^2/\hbar^2-x^2}}{2mEL_\perp^2/\hbar^2-x^2+(m L_\perp/\hbar^2)^2(-V_0 - \alpha x)^2}\Bigg)\nonumber\\
&\equiv \varrho(E,z_0)_{+\alpha x} + \varrho(E,z_0)_{-\alpha x}\,.
\label{Eq:fsDOSint}
\end{align}
The contributions to the density of states from the two bands
\begin{figure}
  \includegraphics[scale=0.35]{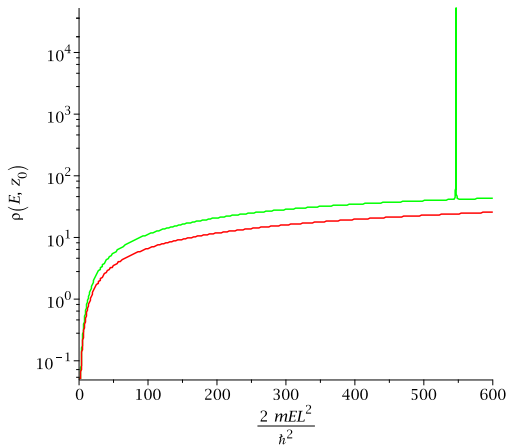}
  \caption{(Colour online) The spin-split free-state density of states at $z=z_0$ in units of $m/(2\pi^2 \hbar^2 L_\perp)$. The green and red curves correspond to, respectively, $\varrho(E,z_0)_{-\alpha x}$ and $\varrho(E,z_0)_{+\alpha x}$ in Eq.~(\ref{Eq:fsDOSint}). At $2mEL_\perp^2/\hbar^2=\beta^2/\eta^2$, $\varrho(E,z_0)_{-\alpha x}$ exhibits a van Hove type singularity.}
  \label{Fig:FSDOSVH}
\end{figure}
are signified by 
the sign of the $\alpha x$ terms in Eq.~(\ref{Eq:fsDOSint}), and correspond to the $b_\pm(E,\beta,\eta)$ terms in the total density of states,
\begin{align}
\varrho(E,z_0)&=\frac{m\Theta(E)}{2\pi^2\hbar^2L_\perp (1-\eta^2)\sqrt{d(E,\beta,\eta)}}\nonumber \\
&\times\Bigg[b_+(E,\beta,\eta)\Bigg(\sqrt{\frac{2mEL_\perp^2}{\hbar^2}}\nonumber \\
&-c_+(E,\beta,\eta)\arctan\Bigg(\frac{\sqrt{2mEL_\perp^2/\hbar^2}}{c_+(E,\beta,\eta)}\Bigg)\Bigg)\nonumber \\
&+b_-(E,\beta,\eta)\Bigg(\sqrt{\frac{2mEL_\perp^2}{\hbar^2}}\nonumber \\
&-c_-(E,\beta,\eta)\arctan\Bigg(\frac{\sqrt{2mEL_\perp^2/\hbar^2}}{c_-(E,\beta,\eta)}\Bigg)\Bigg)\Bigg],\label{Eq:totFSDOS}
\end{align}
where
\begin{align}
b_\pm(E,\beta,\eta)&=\pm\eta\beta+\sqrt{d(E,\beta,\eta)}\,,\nonumber \\
c_\pm(E,\beta,\eta)&=\sqrt{h_\pm(E,\beta,\eta)-2mEL_\perp^2/\hbar^2}\,,\nonumber \\
d(E,\beta,\eta)&=\beta^2+\frac{2mEL_\perp^2}{\hbar^2}(1-\eta^2)\,,\nonumber \\
h_{\pm}(E,\beta,\eta)&=\Bigg(\frac{\sqrt{d(E,\beta,\eta)}\pm\eta\beta}{1-\eta^2}\Bigg)^2.
\end{align}
The spin-split density of states, $\varrho(E,z_0)_{+\alpha x}$ and $\varrho(E,z_0)_{-\alpha x}$ in Eq.~(\ref{Eq:fsDOSint}), is displayed in Fig.~\ref{Fig:FSDOSVH}.
Low-energy contributions to the free-state density of states are dominated by the $\arctan$ terms in Eq.~(\ref{Eq:totFSDOS}) with the three-dimensional $\sqrt{E}$ behaviour appearing for $E\gg E_{\rm min}$. The $\varrho(E,z_0)_{-\alpha x}$ term in Eq.~(\ref{Eq:fsDOSint}) contains a van Hove 
singularity at $2mEL_\perp^2/\hbar^2=\beta^2/\eta^2$, which corresponds to $k_\|=V_0/\alpha$, $k_\perp=0$ and the energy cap on the $E_-$ branch. 
This singularity originates in the $\partial k_\perp/ \partial E \sim 1/k_\perp$ term in Eq.~(\ref{Eq:DOSwffsbs2}). $k_\perp=0$ does not necessarily 
generate a singularity in the density of states due to $|\langle \boldsymbol{x}|\boldsymbol{k}_\|,k_\perp\rangle_{\pm} |^2$ multiplied to $|\partial k_\perp/ \partial E|$ in Eq.~(\ref{Eq:DOSwffsbs2}), with the wavefunction given in Eq.~(\ref{Eq:fswfAP}) vanishing sufficiently quickly.
However, when the bound states cease to exist in the $E_-$ branch, 
the $1/k_\perp$ divergence is no longer compensated, resulting in the van Hove singularity.

\section{\label{sec:level5}Conclusions} 

We have constructed a model Hamiltonian for electrons in a three-dimensional system with an interface, in which the electrons can be trapped by an attractive potential and experience RSOC.
We have analytically obtained the bound-state and free-state wavefunctions, and the 
density 
of states at the location of the interface. We find that the density of states in the interface exhibits two-dimensional behaviour due to the bound states as well as three-dimensional behaviour in the high-energy limit, $E\gg E_{\rm min}=-mV_0^2L_\perp^2/2\hbar^2(1-\eta^2)$. 
The requirement of $\eta^2=(m\alpha L_\perp/\hbar^2)^2\leq1$ 
for a bound state to exist
results in a maximum allowed energy of $\hbar^2V_0^2/2m\alpha^2$ for one of the spin-split bands.  
This restriction impacts the allowed populations of each energy band and enhances the Edelstein and inverse Edelstein effects. Furthermore, the existence of the upper bound in one of the energy bands in the interface leads to a van Hove type singularity in the free-state density of states.
Our results for the density of states are analytical and readily applicable to interfaces and surfaces with RSOC.  

\section{\label{sec:acknowledgements}Acknowledgements}

The research was supported by the Natural Sciences and Engineering Research Council of Canada.

\end{document}